\documentclass[amsmath,amssymb,aps,twocolumn,superscriptaddress,floatfix,longbibliography]{revtex4-2}

\usepackage{bm} 
\usepackage{graphicx} 
\usepackage{comment} 
\usepackage{xcolor} 
\usepackage[normalem]{ulem} 

\begin{document}

\title{An analysis of anomalous particle flow between two correlated systems}

\author{Sirawit Kajonsombat}
\author{Isara Chantesana}
\affiliation{Quantum Computing and Theory Research Centre, Faculty of Science, King Mongkut's University of Technology Thonburi, Bangkok, 10140, Thailand}
\author{Tanapat Deesuwan}
\email[]{tanapat.dee@kmutt.ac.th}
\affiliation{Quantum Computing and Theory Research Centre, Faculty of Science, King Mongkut's University of Technology Thonburi, Bangkok, 10140, Thailand}
\affiliation{Department of Physics, Faculty of Science, King Mongkut's University of Technology Thonburi, Bangkok 10140, Thailand}

\date{\today}

\begin{abstract}
We study the effect of correlation on the direction of particle exchange between local thermal sub-systems where the total system is isolated. Our focus is the situation where both sub-systems have the same temperature but different chemical potentials to eliminate the effect of energy transfer due to the temperature difference. The analysis is done in two limits; in the short time scale where the final state of each sub-system is close to its initial thermal state and in a longer time scale where each sub-system's final state can be arbitrary. The results indicate that the conventional flow of particles from a higher chemical potential to a lower one occurs when the correlation which is quantified by mutual information increases. In contrast, an anomalous flow of particles in the reverse direction has a chance to happen when the correlation goes down. Our findings show that the direction of the particle exchange cannot be predetermined by the chemical potential difference in the presence of correlation. 
\end{abstract}

\maketitle

\section{\label{sec:Intro}{Introduction}}

The concept of correlation is essential to a deeper insight into thermodynamics \cite{bera2017generalized}. For example, it has been found that the direction of the average heat transfer between two sub-systems that forms an isolated system generally cannot be predetermined by their temperature difference when the two sub-systems are non-negligibly correlated. In other words, it is possible for energy to spontaneously move from a colder body to a hotter body in the presence of internal correlation. This anomalous flow of energy has been investigated both theoretically \cite{partovi2008entanglement,jennings2010entanglement,henao2018role,latune2019heat,ma2022anomalous} and experimentally \cite{micadei2019reversing}. Numerical simulations have also been done both on classical computer \cite{gonzalez2020heat} and quantum computer \cite{naghdi2022inducing}. 
The anomalous flow leads to the introduction of the apparent temperature which takes the correlation into account in an attempt to be able to predetermine the direction of energy flow \cite{latune2019apparent}. Moreover, the correlation also plays a critical role in the process of work extraction \cite{vitagliano2018trade,manzano2018optimal,allahverdyan2004maximal,touil2021ergotropy,salvia2022extracting,salvia2023optimal}, can be seen as a resource of work in the resource theory \cite{friis2016energetics,sapienza2019correlations,ng2019resource}, and leads to deeper insights into the concepts of nano and quantum thermal machine \cite{muller2018correlating,holdsworth2022heat,myers2022quantum}, and quantum battery \cite{kamin2020entanglement,liu2021entanglement,shi2022entanglement,garcia2020fluctuations,gyhm2022quantum,arjmandi2022performance}.

In this paper, we analyze the relationship between the change in correlation and the direction of particle flow. Our model is an isolated bipartite system that consists of sub-system $\mathrm{A}$ and $\mathrm{B}$. The sub-systems are allowed to be correlated at an initial time so the total entropy can be written generally as \cite{thomas2006elements}, 
\begin{equation}\label{I_def}
S_\mathrm{AB} = S_\mathrm{A} + S_\mathrm{B} - I_{\mathrm{AB}} \,,
\end{equation}
where $S_\mathrm{AB}$ is total entropy, $S_\mathrm{A}$ and $S_\mathrm{B}$ are marginal entropy of sub-system $\mathrm{A}$ and $\mathrm{B}$ respectively, and $I_{\mathrm{AB}}$ is mutual information that quantifies the correlation between sub-systems. The initial state of each sub-system is assumed to be in its own local thermal equilibrium such that the temperature and chemical potential are locally well-defined.  Our focus is the particle exchange between sub-systems. The analysis is divided into two parts. The first part is done within the limit that the state of each sub-system is close to the thermal equilibrium which is valid in the short time window after the initial time. In the second part, the analysis is extended to the longer time scale where the state of each sub-system can be far from its initial local thermal state. The relative entropy between the final state of each sub-system and its initial thermal state is considered in the same manner as it has been done previously in the study of energy flow \cite{partovi2008entanglement,jennings2010entanglement}. Here, there is no restriction on the final state of the sub-system. 

The paper is organized as follows. The analysis of particle exchange in the presence of correlation within the short time window where sub-system states are close to thermal equilibrium is done in Sec. \ref{sec: 1st result}. In Sec. \ref{sec:2nd result}, an analysis is done in a longer time scale using relative entropy. The conclusion is given in Sec. \ref{sec:Conclusion}. 

\section{\label{sec: 1st result}{Anomalous particle flow near local equilibrium}}

In this section, we investigate the role of mutual information that affects the direction of particle flow between local thermal sub-systems. Thus, the thermodynamic variables such as temperature and chemical potential are locally well-defined with respect to each sub-system. We are interested in the exchange of particles due to the chemical potential difference, and thus, the scenario where both sub-systems share the same temperature but not chemical potential is our focus. 

The system we consider is an isolated system that consists of two sub-systems $\mathrm{A}$ and $\mathrm{B}$. As such, the exchange of energy and particle only occur internally between sub-systems. We further assume that apart from the particle exchange, there is no other form of work exchange such that the energy is only a function of entropy and particle number, $E = E(S,N)$. Since the total system is isolated, the total energy and the total number of particles are conserved, 
\begin{eqnarray}
\delta E = 0 \,, \\
\delta N = 0 \,, 
\end{eqnarray}
where $E$ and $N$ are the total energy and the total particle number respectively. Furthermore, we also impose that the entropy of the total system is also conserved, thus,
\begin{equation}
\delta S_{\mathrm{AB}} = 0 \,, 
\end{equation}
where $S_{\mathrm{AB}}$ is the entropy of the total system. Indeed, the conservation of total entropy may seem to be counter-intuitive from the viewpoint of conventional thermodynamics and the second law of thermodynamics since it is usually stated that total entropy always tends to grow \cite{blundell2010concepts}. However, it will be seen later that such a scenario is now equivalent to an increase in mutual information. 

From Eq.(\ref{I_def}), the change in mutual information reads
\begin{equation}
\delta I_{\mathrm{AB}} = \delta S_{\mathrm{A}} + \delta S_{\mathrm{B}} \,,
\end{equation}
where $S_{\mathrm{A}}$ and $S_{\mathrm{B}}$ are the entropy of sub-system $\mathrm{A}$ and $\mathrm{B}$ respectively. The initial state of the sub-system is assumed to be a thermal state. This implies that the entropy of the sub-system is a thermodynamic function. For example, $S_{\mathrm{A}} = S_{\mathrm{A}}(E_{\mathrm{A}}, N_{\mathrm{A}})$ where $E_{\mathrm{A}}$ is an energy of sub-system $\mathrm{A}$ and $N_{\mathrm{A}}$ is a particle number in sub-system $\mathrm{A}$. The same goes for sub-system $\mathrm{B}$. Thus, the change in entropy is due to the change in energy and particle number. This yields
\begin{eqnarray}\label{delI_master}
 \delta I_\mathrm{AB} &=& \left(\dfrac{\partial{S_\mathrm{A}}}{\partial{N_\mathrm{A}}}\right)_{E_\mathrm{A}} \delta{N_\mathrm{A}} + \left(\dfrac{\partial{S_\mathrm{A}}}{\partial{U_\mathrm{A}}}\right)_{N_\mathrm{A}} \delta{E_\mathrm{A}} 
 \nonumber \\
&& \quad + \left(\dfrac{\partial{S_\mathrm{B}}}{\partial{N_\mathrm{B}}}\right)_{E_\mathrm{B}} \delta{N_\mathrm{B}} + \left(\dfrac{\partial{S_\mathrm{B}}}{\partial{E_\mathrm{B}}}\right)_{N_\mathrm{B}} \delta{E_\mathrm{B}} \nonumber \\
&=& -\big( \beta_\mathrm{A} \mu_\mathrm{A} - \beta_\mathrm{B} \mu_\mathrm{B}  \big) \delta{N_\mathrm{A}} + \big( \beta_\mathrm{A}  - \beta_\mathrm{B} \big) \delta{E_\mathrm{A}} \,,
\end{eqnarray}
where we applied the relations between the thermodynamic variables,
\begin{equation}\label{Legen_1}
\left(\dfrac{\partial S_{\mathrm{A}}}{\partial E_{\mathrm{A}}}\right)_{N_\mathrm{A}} = \beta_{\mathrm{A}} \,,  \qquad 
\left(\dfrac{\partial S_{\mathrm{A}}}{\partial N_{\mathrm{A}}}\right)_{E_\mathrm{A}} = - \beta_{\mathrm{A}} \mu_{\mathrm{A}} \,.
\end{equation}
Here, $\beta_{\mathrm{A}}$ and $\mu_{\mathrm{A}}$ are inverse temperature and chemical potential of sub-system $\mathrm{A}$ respectively. A similar relation was applied for sub-system $\mathrm{B}$. It was also assumed that the interaction between sub-systems does not contribute any extra energy to the whole system such that
%
\begin{eqnarray}
E = E_\mathrm{A} + E_\mathrm{B} \,, \\
N = N_\mathrm{A} + N_\mathrm{B} \,.
\end{eqnarray}
These conditions imply that any change in energy or particle number in one sub-system is compensated by the change in another subsystem. Therefore, $\delta E_\mathrm{A} = -\delta E_\mathrm{B}$ and $\delta N_\mathrm{A} = -\delta N_\mathrm{B}$.

It should be emphasized that Eq.(\ref{delI_master}) is derived as a small fluctuation around the thermal solution. Thus, all terms with higher power of $\delta N$ and $\delta E$ have been neglected. This approximation is valid within a short time scale compared to the dynamical time scale right after both sub-systems are allowed to interact with each other. 

In the limit that both sub-systems have an identical inverse temperature, $\beta_\mathrm{A} = \beta_\mathrm{B} = \beta$ but not chemical potential, Eq.(\ref{delI_master}) is reduced into 
\begin{equation}\label{delI_master_red}
\delta I_\mathrm{AB} = -\beta \big( \mu_\mathrm{A} - \mu_\mathrm{B} \big) \delta N_\mathrm{A} \,.
\end{equation}
Eq.(\ref{delI_master_red}) presents the direct relationship between the change of mutual information and the change in sub-system particle number. 

\subsection{Conventional flow of particles}

In the situation that the mutual information increases, $\delta I_{\mathrm{AB}} > 0$, Eq.(\ref{delI_master_red}) implies
\begin{equation}\label{no_cond_0}
\beta \big( \mu_\mathrm{A} - \mu_\mathrm{B} \big) \delta N_\mathrm{A} < 0 \,.
\end{equation}
If $\mu_\mathrm{A} > \mu_\mathrm{B}$, we immediately obtain that $\delta N_\mathrm{A} < 0$. Similarly, if $\mu_\mathrm{A} < \mu_\mathrm{B}$, then $\delta N_\mathrm{A} > 0$. This demonstrates that when mutual information goes up, there is a flow of particles from a higher chemical potential to a lower one. 

Such a scenario is in line with conventional thermodynamics and the second law, however, without an increase in total entropy. Here, in our setup, mutual information has taken up the role of total entropy in conventional language. Thus, all phenomena that used to be explained by an increase in total entropy are now explained by an increase in mutual information.

\subsection{Anomalous flow of particles}

In contrast to the above situation, when the mutual information decreases, $\delta I_\mathrm{AB} < 0$, then,
\begin{equation}\label{ano_cond_0}
\beta \big( \mu_\mathrm{A} - \mu_\mathrm{B} \big) \delta N_\mathrm{A} > 0 \,.
\end{equation}
The scenario is reversed compared to the previous one. For example, $\delta N_\mathrm{A} < 0$ if $\mu_\mathrm{A} < \mu_\mathrm{B}$. The particle now flows from the part with a \textit{lower} chemical potential to a \textit{higher} one. The occurrence of this manner of flow without the aid of external work is very unlikely in the conventional setting of thermodynamics since the sum of the entropies of all the sub-systems in an isolated system cannot decrease. However, by taking mutual information into account, such a restriction is lifted. It is now possible for the sum of the individual entropies to decrease as long as the mutual information also decreases in such a way that the total entropy of the isolated system does not change. Furthermore, there is an upper bound for mutual information between finite systems. This means that if it is at the highest value initially, the conventional flow will definitely not occur.

To summarise the key point of this section, Ineq. (\ref{no_cond_0}) and (\ref{ano_cond_0}) tell us that one needs to keep in mind that the flow of particles cannot be solely determined by the chemical potential difference in general. The change in correlation which is quantified by mutual information is also needed. It should also be noted that the correlation between sub-systems can be viewed as an extra source of work \cite{bera2017generalized}. Under this viewpoint, the anomalous flow occurs due to the work that has been drawn out from the correlation and does not violate the second law of thermodynamics. 

\section{\label{sec:2nd result}Anomalous particle flow far from equilibrium}

The anomalous flow of heat from a colder body to a hotter body has been investigated in both theory and experiment \cite{partovi2008entanglement,jennings2010entanglement,henao2018role,latune2019heat,micadei2019reversing,gonzalez2020heat,holdsworth2022heat,naghdi2022inducing,bera2017generalized}. It is noteworthy that the mutual information can either be classical or quantum correlation \cite{jennings2010entanglement}.
The crux of the idea is similar to what has been discussed in Sec.\ref{sec: 1st result}. However, unlike the case of small change near equilibrium where the direction of heat transfer absolutely depends on the sign of the mutual information change, when the states involved are not close to equilibrium, the results of the above-mentioned work indicated that one can only say there is a chance that an anomalous heat flow may happen when the mutual information decreases.  

In this section, we would like to analyze the particle flow again, but the exchange is no longer treated as a small fluctuation around the thermal state. The final state of sub-systems after the exchange can be arbitrary. 

The setting that we will use is a bipartite quantum system where each sub-system is initially in its own local thermal state. However, we would like to emphasize that the final results are valid also in the classical setups.

The total Hamiltonian reads
\begin{equation}
\hat H_{\mathrm{tot}} = \hat H_{\mathrm{A}} + \hat H_{\mathrm{B}} + \hat H_{\mathrm{I}} \,,
\end{equation}
where $\hat H_{\mathrm{A}}$ and $\hat H_{\mathrm{B}}$ are the bare Hamiltonian of sub-system $\mathrm{A}$ and $\mathrm{B}$ respectively. $\hat H_{\mathrm{I}}$ is a time-independent interaction Hamiltonian between sub-systems whom we assume to be commuted with the sum of the bare Hamiltonians,
\begin{equation}\label{comm_baretoI}
[\hat H_{\mathrm{A}} + \hat H_{\mathrm{B}}, \hat H_{\mathrm{I}}] = 0 \,.
\end{equation}
Eq.(\ref{comm_baretoI}) ensures that the sum of the bare energies remains constant along the unitary time evolution. As such, the interaction only allows the sub-systems to exchange energy but does not contribute to the total amount of energy of the whole system. Thus, any change in the energy of sub-system $\mathrm{A}$ is compensated by the change in the energy of sub-system $\mathrm{B}$, $\Delta E_\mathrm{A} = -\Delta E_\mathrm{B}$ \cite{micadei2019reversing}. Similarly, we also impose
\begin{equation}\label{comm_NtoI}
[\hat N_\mathrm{tot},\hat H_\mathrm{tot}] = 0 \,,
\end{equation}
where $\hat N_\mathrm{tot} = \hat N_\mathrm{A} + \hat N_\mathrm{B}$ is the total number operator. The operators $\hat N_\mathrm{A}$ and $\hat N_\mathrm{B}$ are the particle number operator of sub-system $\mathrm{A}$ and $\mathrm{B}$ respectively. As a consequence, the sum of the particle number in sub-system $\mathrm{A}$ and $\mathrm{B}$ is conserved, and  $\Delta N_\mathrm{A} = -\Delta N_\mathrm{B}$.

The initial state of the total system is given by the density operator $\hat \rho_\mathrm{AB}(\mathrm{t_i})$ which is not a (grand canonical) thermal state, in general. However, the reduced density operator of each sub-system at the initial time $\mathrm{t_i}$ is. For example, the sub-system $\mathrm{A}$ reduced density operator, $\hat \rho_\mathrm{A}(\mathrm{t_i}) = \mathrm{Tr}_\mathrm{B}[\rho_\mathrm{AB}(\mathrm{t_i})]$, reads
\begin{equation}\label{rhoi_int}
\hat \rho_{\mathrm{A}}(\mathrm{t_i}) = \frac{e^{-\beta_\mathrm{A}( \hat H_\mathrm{A} - \mu_\mathrm{A} \hat N_\mathrm{A})}}{Z_\mathrm{A}} \,,
\end{equation}
where $Z_\mathrm{A}$ is the partition function of sub-system $\mathrm{A}$ at the initial time, and it is given by 
\begin{equation}
Z_\mathrm{A} = \mathrm{Tr}\{ \exp [-\beta_\mathrm{A}( \hat H_\mathrm{A} - \mu_\mathrm{A} \hat N_\mathrm{A})] \} \,.
\end{equation}
In order to have an initial well-defined energy and particle number in each sub-system, we further impose
\begin{equation}
[\hat H_A, \hat N_A] = 0 \,, 
\end{equation}
Note that both $\hat H_\mathrm{A}$ and $\hat N_\mathrm{A}$ are treated as time-independent operators. The time-dependent of the sub-system energy and particle number is due to the time evolution of the density operator. A similar setup is used also for sub-system $B$. We emphasize that the local thermal state of sub-systems $\mathrm{A}$ and $\mathrm{B}$ does not enforce $\hat \rho_\mathrm{AB}(\mathrm{t_i})$ to be a product state of $\hat \rho_{\mathrm{A}}(\mathrm{t_i})$ and $\hat \rho_{\mathrm{B}}(\mathrm{t_i})$. There can be some initial correlation and, therefore, the non-vanishing mutual information between the sub-systems.

The analysis follows what had been done in Refs. \cite{partovi2008entanglement,jennings2010entanglement} for the energy exchange. We used the positivity of the relative entropy to formulate inequalities. The relative entropy is given by \cite{thomas2006elements}
\begin{equation}
S(\rho || \sigma) = \mathrm{Tr}[ \hat \rho \ln ( \hat \rho - \hat \sigma) ]  \geq 0 \,.
\end{equation}
We inspect the relative entropy of the sub-system density operator at the time $\mathrm{t_f}$ with respect to its density operator at the initial time. For sub-system $\mathrm{A}$, it reads
\begin{eqnarray}
&&S[\hat \rho_\mathrm{A}(\mathrm{t_f}) || \hat \rho_\mathrm{A}(\mathrm{t_i}) ] \nonumber \\
&& \qquad  = -\mathrm{Tr}[\hat \rho_\mathrm{A}(\mathrm{t_f}) \ln \hat \rho_\mathrm{A}(\mathrm{t_i})] - S(\hat \rho_\mathrm{A}(\mathrm{t_f})) \nonumber \\
&& \qquad  =  \beta_\mathrm{A} \{ \mathrm{Tr}[\hat \rho_\mathrm{A}(\mathrm{t_f})\big( \hat H_\mathrm{A} - \mu_\mathrm{A} \hat N_\mathrm{A} \big) ]  - \Omega_\mathrm{A} \} - S(\hat \rho_\mathrm{A}(\mathrm{t_f})) \nonumber \\
&& \qquad =  \beta_\mathrm{A} \big( \Delta E_\mathrm{A} - \mu_\mathrm{A} \Delta N_\mathrm{A}\big) - \Delta S_\mathrm{A} \geq 0 \label{A_ineq_1}\,,
\end{eqnarray}
where we substituted the thermal state of $\rho_\mathrm{A}(\mathrm{t_i})$ by Eq.(\ref{rhoi_int}) and replaced the partition function by the grand potential
\begin{eqnarray}
\ln Z_\mathrm{A} &=& -\beta_\mathrm{A} \Omega_\mathrm{A}  \nonumber \\
&=& S(\hat \rho_\mathrm{A}) - \beta_\mathrm{A} \mathrm{Tr}[\hat \rho_\mathrm{A} (\hat H_\mathrm{A} -  \mu_\mathrm{A} \hat N_\mathrm{A})]  \,. 
\end{eqnarray}
Finally, we defined 
\begin{eqnarray}
\Delta E_\mathrm{A} &=& \mathrm{Tr}\{ [\hat \rho_\mathrm{A}(\mathrm{t_f}) - \hat \rho_\mathrm{A}(\mathrm{t_i}) ] \hat H_\mathrm{A}\} \,, \\
\Delta N_\mathrm{A} &=& \mathrm{Tr}\{ [\hat \rho_\mathrm{A}(\mathrm{t_f}) - \hat \rho_\mathrm{A}(\mathrm{t_i}) ] \hat N_\mathrm{A}\} \,, \\
\Delta S_\mathrm{A} &=& S[\hat \rho_\mathrm{A}(\mathrm{t_f})] - S[\hat \rho_\mathrm{A}(\mathrm{t_i})] \,. 
\end{eqnarray}
to be the change of the energy, particle number and entropy of sub-system $\mathrm{A}$ between $\mathrm{t_f}$ and $\mathrm{t_i}$ respectively. A similar inequality can also be found for sub-system $\mathrm{B}$
\begin{eqnarray}
&&S[\hat \rho_\mathrm{B}(\mathrm{t_f}) || \hat \rho_\mathrm{B}(\mathrm{t_i}) ] \nonumber \\
&& \qquad \qquad =  \beta_\mathrm{B} \big( \Delta E_\mathrm{B} - \mu_\mathrm{B} \Delta N_\mathrm{B}\big) - \Delta S_\mathrm{B} \geq 0 \label{B_ineq_1} \,.
\end{eqnarray}
The conservation of the total information $\Delta S_{\mathrm{AB}} = 0$, Eq.(\ref{I_def})  and inequalities (\ref{A_ineq_1}) and (\ref{B_ineq_1}) give
\begin{equation}
(\beta_\mathrm{A} - \beta_\mathrm{B})\Delta E_\mathrm{A} - (\beta_\mathrm{A}\mu_\mathrm{A} - \beta_\mathrm{B}\mu_\mathrm{B})\Delta N_\mathrm{A}  \geq \Delta I_{\mathrm{AB}} \,, 
\end{equation}
where we applied $\Delta E_\mathrm{A} = - \Delta E_\mathrm{B}$ and $\Delta N_\mathrm{A} = - \Delta N_\mathrm{B}$ which is the consequence of Eqs.(\ref{comm_baretoI}) and (\ref{comm_NtoI}). 

If $\mu_\mathrm{A} = \mu_\mathrm{B} = \mu$, then
\begin{equation}\label{delI_ineq_Jenning}
(\beta_\mathrm{A} - \beta_\mathrm{B})\big( \Delta E_\mathrm{A} - \mu\Delta N_\mathrm{A} \big)  \geq \Delta I_{\mathrm{AB}} \,.  
\end{equation}
Eq.(\ref{delI_ineq_Jenning}) coincides with a condition for an energy flow in Ref.\cite{jennings2010entanglement}, however, the energy difference $\Delta E_\mathrm{A}$ is replaced by $\Delta E_\mathrm{A} - \mu\Delta N_\mathrm{A}$. Similar to what has been done in Sec.\ref{sec: 1st result}, the anomalous flow from a colder body to a hotter one can be observed. However, it is not a flow of energy alone but a combination of energy and particle number.

Our focus is the limit that both sub-systems share the same temperature, $\beta_\mathrm{A} = \beta_\mathrm{B} = \beta$ but not a chemical potential in order to study the flow of particle numbers due to the chemical potential difference. An inequality similar to Eq.(\ref{delI_master_red}) is obtained,  
\begin{equation}\label{delI_ineq}
- \beta (\mu_\mathrm{A} - \mu_\mathrm{B})\Delta N_\mathrm{A}  \geq \Delta I_{\mathrm{AB}} \,.  
\end{equation}
Eq.(\ref{delI_ineq}) is no longer a fluctuation around a thermal state. The exchange may be large enough to push the sub-system far out of its initial local thermal state completely. It is straightforward to see that Eq.(\ref{delI_master_red}) can be recovered from Eq.(\ref{delI_ineq}) in the limit that $\rho_\mathrm{A}(\mathrm{t_f})$ is very close to $\rho_\mathrm{A}(\mathrm{t_i})$ such that the relative entropy between them is approximately zero.  It is worth mentioning that $\beta_\mathrm{A}$ and $\mu_\mathrm{A}$ are inverse temperature and chemical potential of the initial state $\rho_\mathrm{A}(\mathrm{t_i})$ which is locally thermal. $\rho_\mathrm{A}(\mathrm{t_f})$ is not necessarily thermal and does not possess well-defined thermodynamical parameters. 

The analysis of the particle exchange between sub-systems then follows what has been done in Sec. \ref{sec: 1st result} and similarly to Ref. \cite{partovi2008entanglement}. The particle exchange between sub-systems in the case that $\Delta I_\mathrm{AB} = 0$ and $\mu_\mathrm{A} = \mu_\mathrm{B} = \mu$ is considered reversible.   

If $\Delta I_\mathrm{AB} > 0$, $\Delta N_\mathrm{A}$ is negative when $\mu_\mathrm{A} > \mu_\mathrm{B}$. This is the conventional flow of particles from a higher chemical potential to a lower one. This scenario is most promoted by the \textit{Stosszahlansatz} where initially the total density operator is a product of its sub-systems density operator, $\rho_\mathrm{AB} = \rho_\mathrm{A} \otimes \rho_\mathrm{B}$. Thus, the initial mutual information is zero, and the time evolution can only increase mutual information. In general situations, the conventional flow is caused by an initial state with a small correlation between sub-systems. 

The anomalous flow of particles from a lower chemical potential to a higher one has a chance to occur when $\Delta I_\mathrm{AB} < 0$, however, it is not definite. We need
\begin{equation}\label{ano_cond}
0 > - \beta (\mu_\mathrm{A} - \mu_\mathrm{B})\Delta N_\mathrm{A}  \geq \Delta I_{\mathrm{AB}} \,,
\end{equation}
which is necessary to have $\Delta I_\mathrm{AB} < 0$ but a decrease in mutual information alone does not guarantee an anomalous flow. This is a major difference compared to the inequality (\ref{ano_cond_0}) where an anomalous flow requires no further condition except $\delta I_\mathrm{AB} < 0$. Due to the exchange of particles no longer keeping both sub-systems in the thermal state, the dynamics becomes more complex such that it can not be predetermined by only a few parameters given by thermodynamics.

\section{\label{sec:Conclusion}Conclusion}

We studied the flow of particles between two local thermal sub-systems in the presence of correlation. Our focus was the case where both sub-systems have the same temperature but different chemical potentials in order to see the particle flow as a result of the chemical potential difference and without the interference of energy transfer due to the temperature difference. 

The analysis was done in two limits: 1. in a short time window where the sub-system states are close to their initial thermal states and 2. in a general case of a longer time window where there is no restriction on the sub-system final states. In both cases, the correlation plays a crucial role in the direction of the particle flow. 

The conventional flow of particles from a higher chemical potential to a lower one is observed when the mutual information between sub-systems increases. The situation is similar to the case of energy flow where energy flows from a higher temperature to a lower one when mutual information between sub-systems is going up \cite{partovi2008entanglement,jennings2010entanglement}. This implies that the conventional phenomena predicted by thermodynamics are recovered when there is an increase in mutual information.  

In contrast, an anomalous flow of particles from a lower chemical potential to a higher one has a chance to happen when there is a decrease in mutual information. The scenario is again identical to the case of energy flow \cite{partovi2008entanglement,jennings2010entanglement}. Note that, in a special case where both systems shift very little from their initial equilibrium and the mutual information decreases, our result indicates that the anomalous particle flow must definitely occur. These phenomena have not been previously considered in the thermodynamic analysis as a result of the correlation between sub-systems being ignored. 

Our theoretical results show that correlation plays a crucial role in particle transfer between two systems as much as in energy transfer. This should lead to a new understanding of physical phenomena that involves particle exchange. Thus, it would be very interesting to check the validity of our results by experiments.

\begin{acknowledgements}
This research has received funding support from the NSRF via the Program Management Unit for Human Resources \& Institutional Development, Research and Innovation [grant number B05F650024] and from National Science and Technology Development Agency [New ST Researcher Supporting Grant 2021]. The authors acknowledge King Mongkut’s University of Technology Thonburi for funding Postdoctoral Fellowship to IC.

\end{acknowledgements}

\bibliography{bibliography}

\end{document}